\documentclass[10pt,a4paper]{article}
\usepackage[T1]{fontenc}
\usepackage{inputenc}
\usepackage[english]{babel}
\usepackage[nohead,includefoot,margin=2cm]{geometry}
\usepackage[compact,raggedright,small]{titlesec}
\usepackage[affil-it]{authblk}
\usepackage[runin]{abstract}
\usepackage{multicol}
\usepackage{blindtext}
\usepackage{graphicx}
\usepackage{caption}

\newcommand*{\pacsname}{PACS numbers}

\addto{\Affilfont}{\small}

\title{\large\bfseries Modeling KIC 10684673 and KIC 12216817 as Single Pulsating Variables}
\date{}
\author[1]{Garrison Turner \thanks{gturner0040@kctcs.edu}}
\author[1]{Austin Maynard \thanks{amaynard0034@kctcs.edu}}
\affil[1]{Dept. of Physics and Astronomy, Big Sandy Community and Technical College, Prestonsburg, KY 41653}

\begin{document}
  \maketitle

  \begin{abstract}
    The raw light curves of both KIC 10684673 and KIC 12216817 show variability. Both are listed in the Kepler Eclipsing Binary Catalog (hereafter KEBC), however both are flagged as uncertain in nature. In the present study we show their light curves can be modeled by considering each target as single, multi-modal $\delta$ Scuti pulsator. While this does not exclude the possibility of eclipsing systems, we argue, while spectroscopy on the systems is still lacking, the $\delta$ Scuti model is a simpler explanation and therefore more probable.\\
  \end{abstract}
  \begin{multicols}{2}
  \section{Introduction}
While the original Kepler mission was in operation, it continually monitored over 150,000 stars in a region of sky containing parts of the Cygnus and Lyre constellations. With unprecedented photometric precision, a great amount of work has been done to categorize, model, and determine the nature of the many new variables discovered (for an overview of mission characteristics and objectives, see Borucki (2008); for an overview of the Kepler Asteroseismology group, see Gilliland (2010). Two categories which have had several members added as a result of the mission are eclipsing binaries and intrinsic pulsating variables. However, as with any rapid increase of such categories, the potential for initial misclassification of a system is possible. Both targets are listed in the KEBC (10684673 with a period of 0.1925563 days and 12216817 with a period of 0.2462731 days; Kirk et al. 2016), however both are flagged as uncertain in nature. The light curves of both targets are shown in Figure 1. While both light curves show variability, they do not share characteristics of classical eclipsing binaries, such as Algol type systems. For example, there are no obvious eclipse features. Rather, the structures in the light curve seem to be a superposition of several sinusoidal signals. As will be shown below, the frequencies and amplitudes of all the dominant frequencies in each target are consistent with δ Scuti-type pulsations. One specific class of eclipsing binaries, W Uma eclipsing binaries, have the same depth for both the primary and secondary eclipse, and can therefore produce, under the right circumstances, a sinusoidal signal. With the presence of the additional signals, there are two logical possibilities. Either the targets could be W Uma systems, in which one of the components is also pulsating, or the light curve can be interpreted as a classic single $\delta$ Scuti pulsating in multiple modes.\\
	
W Uma eclipsing binaries are characterized by near equal minima, variability as low as a few tenths of a magnitude, and short periods of about 0.25 to about 1.0 days. Because the periods are short, most W Uma variables are assumed to be near contact or overcontact binaries. W Uma systems can range from A through K spectral types. See Binnendijk (1970) and Li et al. (2007) for an overview of W Uma systems.\\

\begin{figure*}
\centering
   \includegraphics[width=17cm]{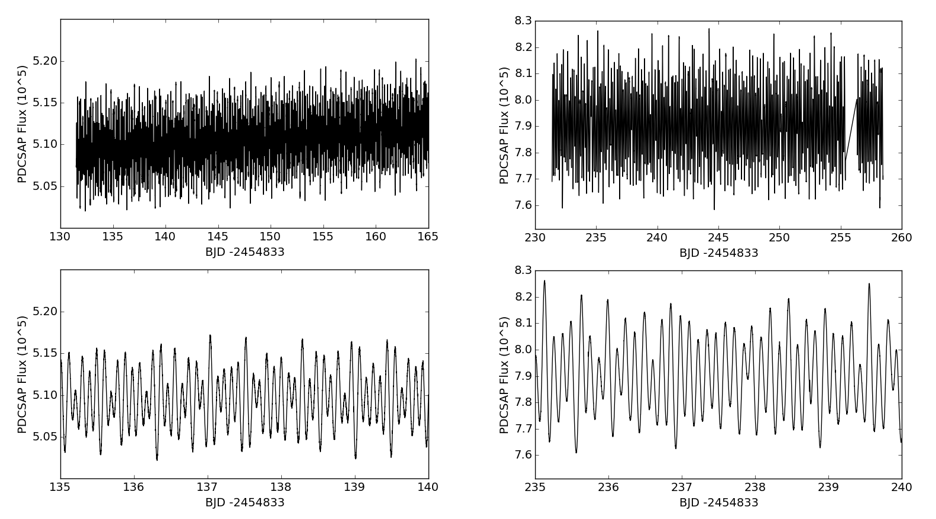}
     \caption{Shown are the light curves for KIC 10684673 and KIC 12216817. The top panels show the entire SC light curves for both targets; KIC 10684673 the left panel and KIC 12216817 the right. The bottom panels show subsections of the light curve, displaying the variability of each target.}
     \label{<Figure 1>}
\end{figure*}

\section{Observations}
In this study, we use exclusively the one short-cadence (SC) data set obtained by the Kepler Space Telescope and retrieved from the MAST archive(Aura 2015). SC data sets are approximately one month in duration (rather than a full three-month quarter) yet with the shortened image collection and integration times, data sets regularly have more than 40,000 data points (as opposed to ~7,000 data points for each long cadence data set). Because the SC data sets have such short time intervals, they are the most useful for studying $\delta$ Scuti-type variability. Both targets have one SC data set publicly available. Kepler began SC data collection for KIC 10684673 at 12:15:49 on 2009-05-13 and ended at 11:47:11 2009-06-15 (taken during Quarter 1 of observations). The start and end collection times for KIC 12216817 are 20:38:32 2009-08-20 and 23:23:42 2009-09-23, respectively (taken during Quarter 2 of the observational cycle). The data set for KIC 10684673 had a total of 49,030 usable datapoints, while KIC 12216817 had a total of 38271, each after the removal of the outliers. Sample PDCSAP flux light curves (after the removal of the Null data points when the telescope was offline) are shown in Figure 1. As can be seen from Figure 1, no obvious eclipse features are prominent in the light curves of either target. The KEBC gives an orbital period of 0.1925563 days for KIC 10684673, and an orbital period of 0.2462731 days for KIC 1226817. Both targets, while in the catalog, carry an UNC flag for uncertain whether the object is an eclipsing binary.\\

\begin{center}
  \includegraphics[height=60mm]{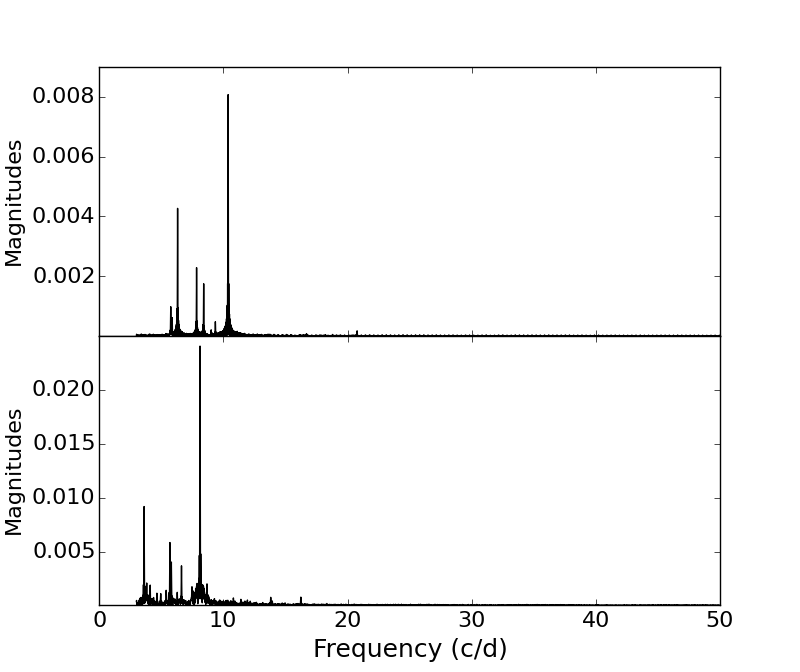}
  \captionof{figure}{Shown are the power spectra for KIC 10684673 and KIC 12216817, top to bottom respectively. The x-axis shows the frequency in (c/d) while the y-axis gives the amplitude results for each frequency in magnitudes.}
\end{center}
\section{Analysis}
To prepare the data for analysis, the first step taken involved removing any obvious outlier data points. Secondly, the fluxes were converted to a magnitude scale via the equation\begin{equation}
M = 2.5log(F)
\end{equation}
where M is the magnitude, and F is the flux of each individual PDCSAP measurement. For convenience, the conventional minus sign was omitted, as we are only interested in $\Delta$M; the minus sign is unnecessary and this equation preserves the orientation of the data as presented in Figure 1. Once converted to the magnitude scale, the data were frequency-searched using the Period04 software package (Lenz and Breger 2005) between the values of 3 c/d and 50 c/d, the typical range of frequencies for $\delta$ Scuti variables. The power spectra for both targets are shown in Figure 2 in terms of the amplitude in magnitudes. As can be seen, both targets have several signals in their respective power spectra. To model the data, only frequencies with a signal-to-noise ratio of $\geq$ 5 were used to save on computational time. Table 1 gives the values of the frequencies used for the modeling and their associated uncertainties.\\
\begin{figure*}
\centering
   \includegraphics[width=17cm]{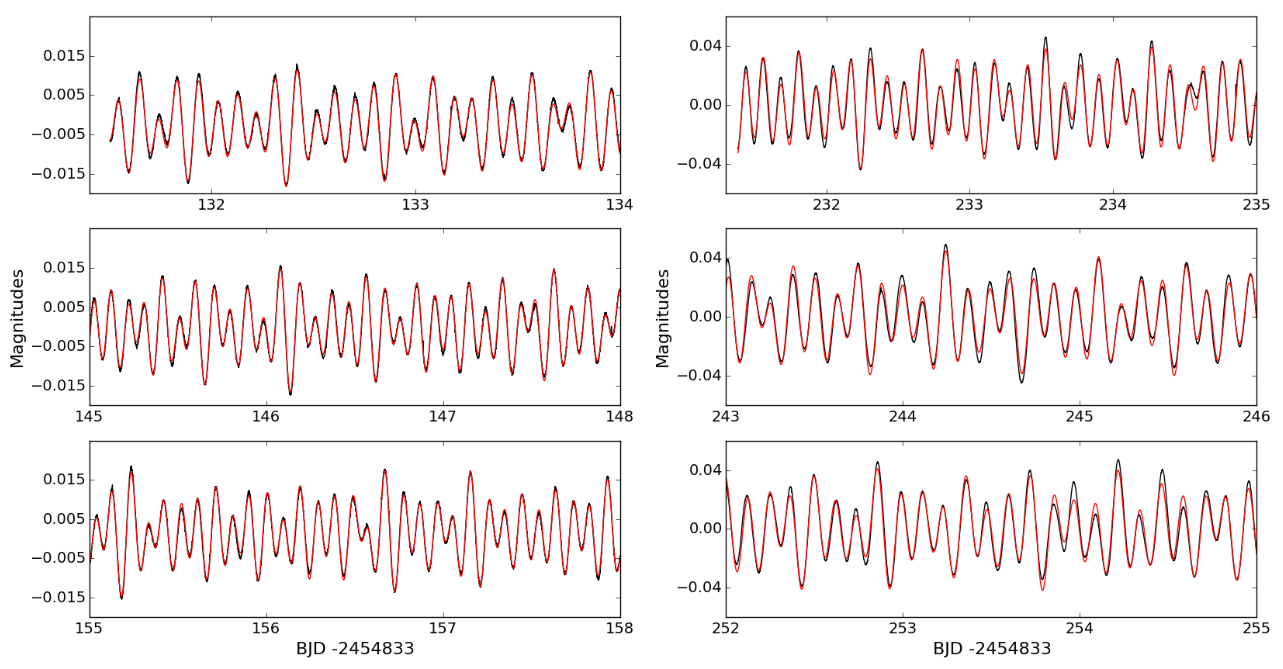}
     \caption{Shown are representative portions of each light curve with the results of the fit from the $\chi^{2}$ fit. The left panels show KIC 10684673 and the right panels show KIC 12216817. The Kepler data is in black, while the red is the result of each fit. Each plot shows a small representative sample of the fit at various intervals throughout each data set to show the agreement of the fit over the entirety of each data set.}
     \label{<Figure 3>}
\end{figure*} 

\begin{table*}[t]
	\centering
	\begin{tabular}{|c|c|c|c|c|}
	\hline 
 	& KIC 1068467 &  &  KIC 12216817 &  \\ 
	\hline 
	m & Frequency (c/d) & $\sigma_{F}$ (c/d) & Frequency (c/d) & $\sigma_{F}$ (c/d) \\ 
	\hline 
	1 & 10.3870092 & 0.0023731 & 8.12105499 & 0.00470000 \\ 
	\hline 
	2 & 6.31430081 & 0.00235175 & 3.61637202 & 0.00470000 \\ 
	\hline 
	3 & 7.84305938 & 0.00233767 & 5.69616625 & 0.00470000 \\ 
	\hline 
	4 & 8.41485091 & 0.00467293 & 5.81058261 & 0.00470000 \\ 
	\hline 
	5 & 5.76341027 & 0.00468387 & 6.62810598 & 0.00940000 \\ 
	\hline 
	\end{tabular}
	\caption{Shown are the frequencies and uncertainties determined from the frequency search for both targets using the Period04 software package.}
	\label{Table 1.}
\end{table*} 

\begin{table*}[t]
	\centering
	\begin{tabular}{|c|c|c|c|c|c|c|c|c|}
    \hline 
     & KIC 10684673 &  &  &  & KIC 12216817 &  &  &  \\ 
    \hline 
    m & Amplitude (Mag) & $\sigma_{A}$ (Mag) & $\phi$ & $\sigma_{\phi}$ & Amplitude (Mag) & $\sigma_{A}$ (Mag) & $\phi$ & $\sigma_{\phi}$ \\ 
    \hline 
    1 & 0.00809 & 0.00001 & 5.3402 & 0.0002 & 0.02403 & 0.00003 & 4.6362 &  0.0001 \\ 
    \hline 
2 & 0.00429 & 0.00003 & 6.2001 & 0.0004 & 0.00913 & 0.00003 & 5.7360 & 0.0005 \\ 
    \hline 
3 & 0.00229 & 0.00002 & 3.8398 & 0.0007 & 0.00580 & 0.00003 & 6.0895 & 0.0008 \\ 
    \hline 
4 & 0.00177 & 0.00003 & 5.9501 & 0.0009 & 0.00392 & 0.00003 & 0.9554 & 0.0011 \\ 
    \hline 
5 & 0.00101 & 0.00005 & 1.9686 & 0.0015 & 0.00366 & 0.00003 & 2.7192 & 0.0012 \\ 
    \hline 
 &  &  &  &  &  &  &  &  \\ 
    \hline 
<$\chi^{2}$> & 7.09 &  &  &  & 16.15 &  &  &  \\ 
    \hline 
    \end{tabular} 
    \caption{Shown are the results of the fitting procedure and the final value of the reduced $\chi^{2}$ calculation.}
    \label{Table 2}
\end{table*}

\section{Modeling}
To model the data, we performed a fitting routine in Period04 using the frequencies of Table 1. From the fit, the amplitudes and phases were determined (along with their uncertainties). These parameters were then taken as the initial estimates for our model which was then inserted into a reduced $\chi^{2}$ calculation, as outlined in Bevington and Robinson (2003). The reduced $\chi^{2}$ can be written as \begin{equation}
<\chi^{2}>=\frac{1}{N}\sum{\left(\frac{y_{i}-y(t_{i})}{\sigma_{i}}\right)}
^{2}
\end{equation}
where N is the number of data points, y$_{i}$ represents the observed magnitudes associated with each observation time, $\sigma_{i}$ represents the uncertainty associated with the measurements, and y(t$_{i}$) represents the value of the model at each observation time. Each PDCSAP flux measurement comes with an associated uncertainty, which must be converted to the uncertainties on the magnitudes. To convert the uncertainties from fluxes to magnitudes, we use the usual procedure
\begin{equation}
\sigma_{m}=\sigma_{F}\sqrt{\left(\frac{\partial M}{\partial F}\right)^{2}}\approx \frac{2.5\sigma_{F}}{2.3F}
\end{equation}
where $\sigma_{m}$ and $\sigma_{F}$ are the uncertainties on the magnitudes and fluxes, respectively. The model y(t$_{i}$) in the reduced $\chi^{2}$ calculation are given by\begin{equation}
y(t_{i})=\sum_{m=1}^{n}A_{m}sin(\omega_{m}t_{i}+\phi_{m})
\end{equation}
where $A_{m}$, $\omega_{m}$, and $\phi_{m}$ are the amplitude, angular frequency, and phase of each sine wave given and $n$ is the total number of frequencies for each object used in the study (coincidentally $n$ = 5 in both cases). It should be noted that in the case of KIC 10684673, there is a definite systematic and positive trend in the data. The entire data set was fit with a linear function to determine the slope and intercept of the trend. Finally, it is customary to define the reduced $\chi^{2}$ function as the original $\chi^{2}$ function divided by the difference between the number of data points and the number of degrees of freedom. However, because the number of data points for both target are $N \simeq$ 40,000, and the number of parameters (p) being fit for both targets are p < 10, the difference becomes $N-p\approx N$.\\
\indent The model was refined by running a nonlinear least-squares fit using a Marquardt-Levenburg algorithm. The routine allowed the amplitudes and phases to vary while holding the frequencies of Table 1 fixed until the set of amplitudes and phases which minimized the reduced $\chi^{2}$ were found. In the case of KIC 10684673, the linear trend was also incorporated into the routine. The values of the parameters (and associated uncertainties) obtained from the fitting procedure are given in Table 2, as is the overall value for the reduced $\chi^{2}$ calculation. The results of the fitting for both targets are shown in Figure 3. This shows representative samples of the fit over the entire SC data set for both targets in order to show the quality of the fit with the data.\\
\indent Ideally, the expectation value of the reduced $\chi^{2}$ for a very good fit would be very close to one (Bevington and Robinson, 2003). As can be seen from Table 2, our values are somewhat high. However, this can be explained by the extreme smallness of the photometric uncertainties associated with each measurement. They are typically about 3 or more orders of magnitude smaller than the measurements. This being the case, even though our reduced $\chi^{2}$ values seem a little high for typical photometric measurements, Kepler's precision is very much atypical. Therefore we conclude, given how well the data is replicated by the model, that the fit is satisfactory, especially since the model replicates every major feature throughout the entire data set. It is also at this point that we can feel confident, given the quality of the models, that the frequencies reported in Table 1 are genuine signals and not spurious.\\

\section{Conclusion}
In this study we have demonstrated that the SC light curves of KIC 10684673 and KIC 12216817 can both be modeled as single $\delta$ Scuti-type pulsating systems pulsating in several modes. Both are included in the KEBC, however both are flagged as uncertain in nature. While our analysis does not completely rule out the possibility of either of these systems being eclipsing binaries, this study gives strong support for the hypothesis that all of the variation in each case can be explained by a single, pulsating object. To completely determine the nature of the system, spectroscopy would need to be performed on both targets.\\
  \end{multicols}
  
\end{document}